\documentclass{article}

\usepackage[a4paper,top=2cm,bottom=2cm,left=2cm,right=1cm]{geometry}
\usepackage{amsmath,mathtools,blkarray,bm}
\usepackage{dcolumn}
\usepackage{listings}
\usepackage{xcolor}
\usepackage{hyperref}
\usepackage{graphicx}
\usepackage{authblk}
\usepackage[width=.75\textwidth]{caption}

\def\t{{\bar t}}
\def\Mtt{m_{t\bar{t}}}
\def\PTavt{p_{\mathrm{T,avt}}}
\def\Yavt{y_{\mathrm{avt}}}
\def\Ytt{y_{t\bar{t}}}
\def\as{\alpha_s}

\begin{document}
\date{}

\title{{\sc fastNLO} tables for NNLO top-quark pair differential distributions}

\author[1]{Micha\l\  Czakon}
\author[2]{David Heymes}
\author[2]{Alexander Mitov}

\affil[1]{{\small Institut f\"ur Theoretische Teilchenphysik und Kosmologie,
RWTH Aachen University, D-52056 Aachen, Germany}}
\affil[2]{{\small Cavendish Laboratory, University of Cambridge, Cambridge CB3 0HE, UK}}

\maketitle

\center{\it Preprint numbers: Cavendish-HEP-17/08, TTK-17-10}\vskip 10mm

\abstract{We release {\sc fastNLO} tables with NNLO QCD top-quark pair differential distributions corresponding to 8 TeV ATLAS \cite{Aad:2015mbv} and CMS \cite{Khachatryan:2015oqa} measurements. This is the first time {\sc fastNLO} tables with NNLO QCD accuracy have been made publicly available. The tables are indispensable in pdf fits and allow, for the first time, very fast calculation of differential distributions with any pdf set and for different values of $\as$. The numerical accuracy of the resulting differential distributions is high and comparable to the accuracy of all publicly available NNLO top-quark differential calculations. We intend to keep producing tables corresponding to existing and future LHC measurements at various collider energies.}

\section{Introduction}

Calculations of LHC processes at next-to-next-to-leading order (NNLO) in perturbative QCD are numerically challenging and at present require computing times of $\mathcal{O}(10^4-10^5)$ CPU hours for a single calculation. Typically, calculations are performed for a pre-decided specific set of bin(s), values of relevant parameters (like masses and coupling constants) and parton distribution functions (pdf). In the following we will refer to such calculations as {\it direct}. Once direct results have been produced, typically in the form of binned histograms, they cannot be changed any more. In other words, if one wishes to perform the same calculation but for different values of $\as$, pdf, mass or set of bins -- even for the same distribution -- one has to repeat the calculation from scratch. Alternative flexible formats for storing results that allow one to reuse existing calculations in future applications are, therefore, highly desirable. For the extraction of pdfs from LHC data where up to $\mathcal{O}(10^5)$ recalculations of the hadronic cross section for different pdfs are required, the speed of the calculation becomes a major additional requirement.

A number of formats with different levels of flexibility have been proposed and are commonly used in NLO calculations. The $n$-tuples event-file format for (weighted) partonic NLO calculations \cite{Bern:2013zja} stores all the information for every partonic event/counterevent. This format allows one to reasonably fast recompute arbitrary observables and modify parameters like the renormalization and factorization scales, pdf's and $\as$. While the $n$-tuple format provides high level of flexibility, it has one disadvantage: event-files tend to be very large with sizes approaching $\mathcal{O}(1)$ TB (depending on the process). This disadvantage becomes more significant at NNLO due to the larger number of events/counterevents that need to be stored for numerically stable results. A first study of the size of the $n$-tuples event-file format for NNLO computations in $e^+e^-$ collisions has been performed in ref.~\cite{Heinrich:2016jad}.

An alternative approach to storing and recomputing NLO calculations is offered by the {\sc fastNLO} \cite{Kluge:2006xs,Wobisch:2011ij,Britzger:2012bs} and {\sc APPLgrid}~\cite{Carli:2010rw} formats. This approach produces and stores for later use an accurate interpolation of the partonic cross-section in a pdf and $\alpha_s$ independent way. This approach enables the extremely fast recalculation of fixed observables with different pdf sets, values of $\as$ and, possibly, factorization and renormalization scales. These interpolation formats are particularly useful for extracting pdfs from data. Importantly, the interpolation approach can be simply extended to NNLO computations and a {\sc fastNLO} interface to event generators at NNLO is publicly available. In contrast to storing event-files, the required disk space for {\sc fastNLO} tables, even at NNLO, does not exceed $\mathcal{O}(100)$ MB and is a convenient way of storing and distributing differential results at NNLO. A drawback of this approach is that one cannot change the distributions or bins.

As a first application of {\sc fastNLO} at NNLO in this work we produce {\sc fastNLO} tables for four one-dimensional top-quark pair differential distributions for LHC at 8 TeV. These same  distributions were computed in refs.~\cite{Czakon:2016dgf,Czakon:2015owf} as binned histograms and for specific pdf sets. To produce the tables we have straightforwardly interfaced the {\sc fastNLO} library to an in-house Monte Carlo (MC) generator which implements the {\sc Stripper} approach for NNLO calculations \cite{Czakon:2010td,Czakon:2011ve,Czakon:2014oma}. The table files are publicly available \cite{web-tables} and can be used, among others, for pdf extractions \cite{Czakon:2016olj}, strong coupling variation and determination and new physics searches.

\section{{\sc fastNLO} tables for top-quark pair production at 8 TeV}

The binning of the produced tables corresponds to the common binning used by ATLAS \cite{Aad:2015mbv} and CMS \cite{Khachatryan:2015oqa} in the lepton plus jets LHC 8 TeV measurements. Tables are produced for the following four distributions: invariant mass of the top-quark pair $\Mtt$, transverse momentum of the averaged top/antitop quark $\PTavt$, rapidity of the average top/antitop quark $\Yavt$ and rapidity of the top-quark pair $\Ytt$. Following ref.~\cite{Czakon:2016dgf}, the renormalization scale $\mu_{\mathrm{R}}$ and factorization scale $\mu_{\mathrm{F}}$ have been set to $m_{T}/2$ for the transverse momentum distribution of the average top/antitop quark and to $H_T/4$ for all other distributions. 

We would like to emphasize that our tables contain only the central values for $\mu_{\mathrm{F,R}}$ and cannot be used for performing scale variation or for changing the functional form of the two scales. The provided information should be more than adequate for most applications, like pdf fitting, where {\sc fastNLO} tables are indispensable. If an estimate of the scale error for calculations performed with our {\sc fastNLO} tables is desired, one could construct such an estimate from the direct calculation in ref.~\cite{Czakon:2016dgf} which is available in electronic form.
\footnote{Since results for several pdf sets are readily available in electronic form \cite{Czakon:2016dgf}, one could easily verify the extend to which the scale variation for each distribution and each bin is pdf-independent.}

A summary of the produced {\sc fastNLO} tables is given in table~\ref{tab:tablesummary}. Their calculation was performed at the University of Cambridge's Darwin cluster \cite{Darwin}. The computing time needed to produce the tables did not significantly exceed the time it took to perform our previous direct calculation \cite{Czakon:2016dgf}. For reasons of computational efficiency
\footnote{We hope that for future {\sc fastNLO} table releases we will be able to provide the results for each distribution within a single table.}
two tables instead of one are provided for the average top/antitop rapidity distribution (each one of the other three distributions is contained within a single table). The full result for $\Yavt$ is obtained after summing the output of the two tables for each perturbative order. The {\sc fastNLO} software and instructions for using {\sc fastNLO} tables can be found on the {\sc fastNLO} website~\cite{fastNLO}. An example for their usage is given in appendix \ref{sec:appendix}.  
\begin{table}[h]
  \begin{center}
    \renewcommand{\arraystretch}{1.5}
    \begin{tabular}{llcc}
      \hline
      Observable & Binning & $\mu_{\mathrm{F}}=\mu_{\mathrm{R}}$& number of files 
      \\ \hline
      $m_{t\bar{t}}\,[\mathrm{GeV}]$ & $\{345,\,400,\,470,\,550,\,650,\,800,\,1100,\,1600\} $ &  $H_T/4$ & 1
      \\
      $p_{\mathrm{T,avt}}\,[\mathrm{GeV}]$ & $\{0,\, 60,\, 100,\, 150,\, 200,\, 260,\, 320,\, 400,\, 500\} $ &  $m_T/2$ & 1 \\
      $y_{\mathrm{avt}}$ & $\{-2.5,\,-1.6,\,-1.2,\,-0.8,\,-0.4,\,0.0,\,0.4,\,0.8,\,1.2,\,1.6,\,2.5
      \} $ &  $H_T/4$ & 2\\
      $y_{t\bar{t}}$ & $\{-2.5,\,-1.3,\,-0.9,\,-0.6,\,-0.3,\,0.0,\,0.3,\,0.6,\,0.9,\,1.3,\,2.5
\} $ &  $H_T/4$ & 1 \\
      \hline
    \end{tabular}
  \end{center}
  \caption{\label{tab:tablesummary} Summary of the {\sc fastNLO} tables provided with this work. Note that the $\Yavt$ distribution is split in two tables whose contributions need to be added.}
\end{table}

\section{Quality of the {\sc fastNLO} tables}\label{sec:quality}

Presently, the {\sc fastNLO} table format does not have a facility for estimating Monte Carlo errors. To control the quality of our tables we perform two tests: we estimate the interpolation error of the tables as well as the accuracy of the distributions by comparing predictions derived from the tables with three independent calculations. 

\begin{figure}[t]
\centering
\hspace{0mm} 
\includegraphics[page=1,width=0.40\textwidth]{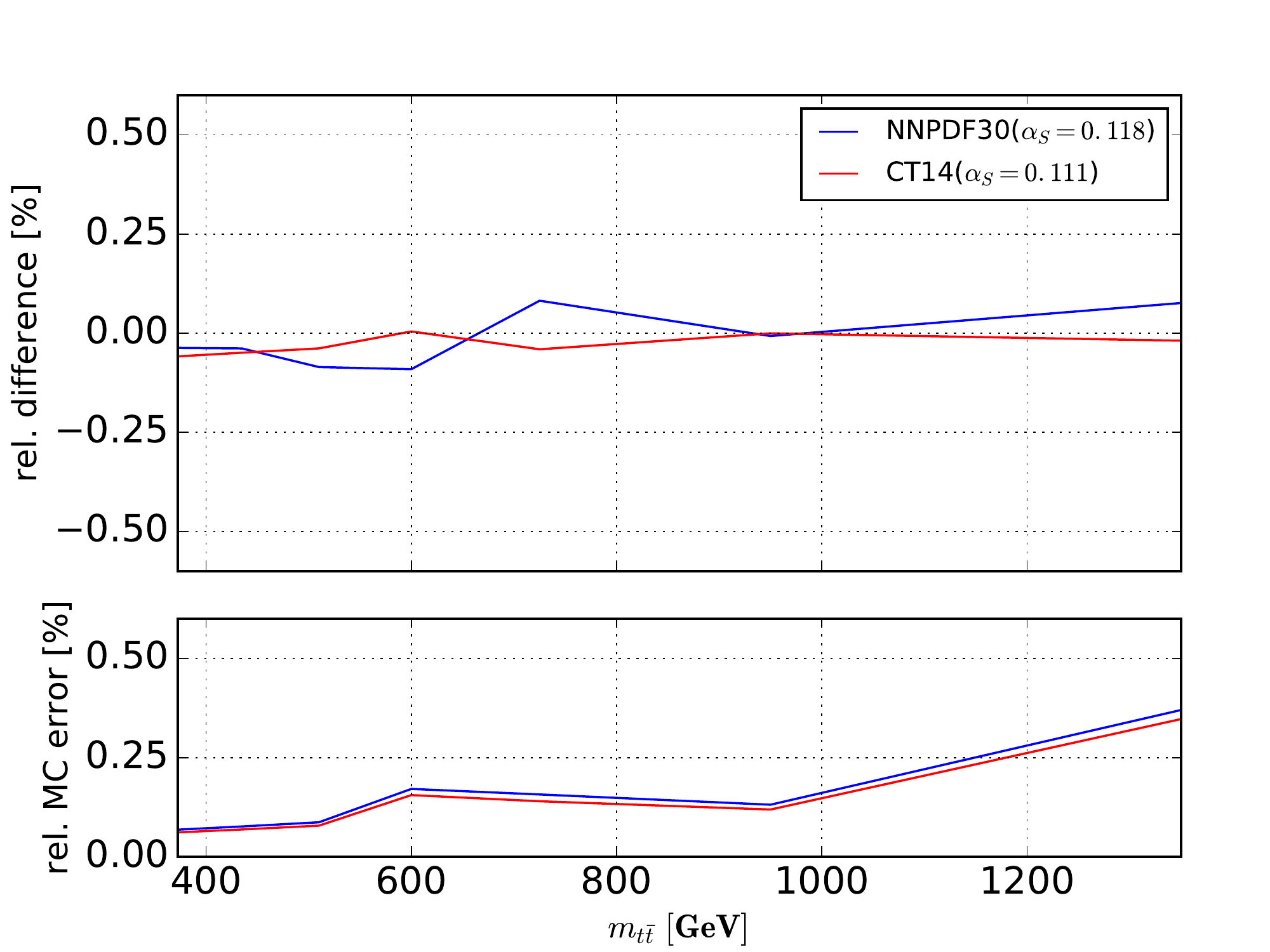}
\includegraphics[page=2,width=0.40\textwidth]{plots/interpolation.pdf}
\includegraphics[page=3,width=0.40\textwidth]{plots/interpolation.pdf}
\includegraphics[page=4,width=0.40\textwidth]{plots/interpolation.pdf}
\caption{Interpolation error (top panels) of our {\sc fastNLO} tables. See sec.~\ref{sec:quality} for details.}
\label{fig:interpolation}
\end{figure}

To check the interpolation error inside our NNLO {\sc fastNLO} tables in fig.~\ref{fig:interpolation} we compare the four differential distributions derived from the tables with direct NNLO calculations that were performed simultaneously with the tables, using exactly the same set of partonic events. This way, the same information from each generated partonic event is passed to the table {\it and} to the histogram, ensuring the calculations of the table and the direct histogram are fully correlated. The final table is then convoluted with the same pdf set used in the direct histogram and the two are compared in fig.~\ref{fig:interpolation}. To ensure pdf-independence of the test, we simultaneously compute two histograms that are based on different pdf sets. We have used NNPDF30~\cite{Ball:2014uwa} and CT14~\cite{Dulat:2015mca} which were chosen to have different values of the strong coupling: $\alpha_s(m_Z) = 0.118$ and $0.111$, respectively. We interpret the relative difference between the table-based and direct calculation for a given pdf set as due to table interpolation. The comparison for all four distributions can be found in the upper panels of fig.~\ref{fig:interpolation}. The interpolation error is about $1$ per mille for all bins for all four distributions and is almost independent of the pdf set. This error is negligible since it is much smaller than the Monte Carlo error of the direct NNLO calculation, which can be found in the lower panels of fig.~\ref{fig:interpolation}.

\begin{figure}[t]
\centering
\hspace{0mm} 
\includegraphics[page=1,width=0.40\textwidth]{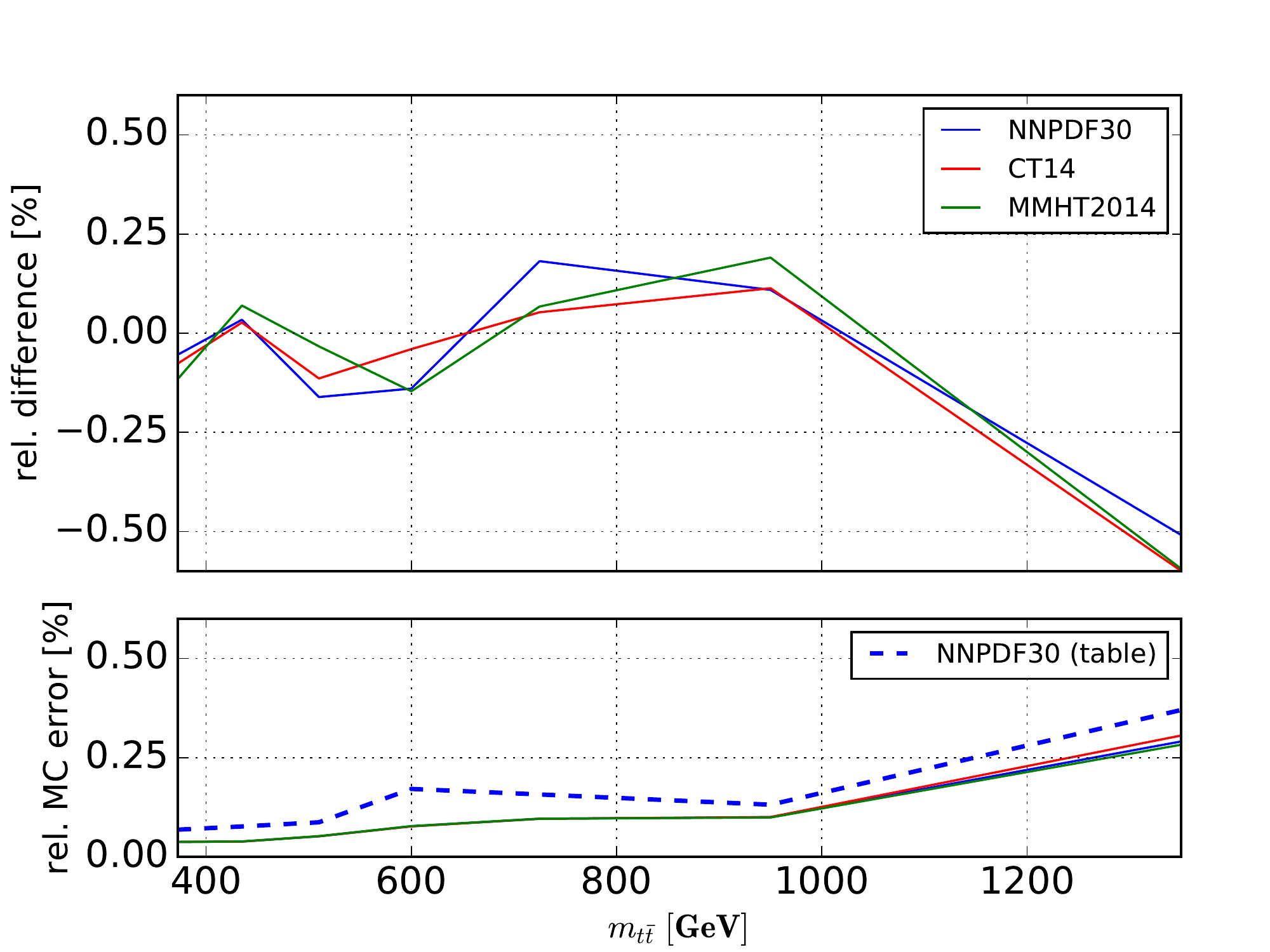}
\includegraphics[page=2,width=0.40\textwidth]{plots/accuracy.pdf}
\includegraphics[page=3,width=0.40\textwidth]{plots/accuracy.pdf}
\includegraphics[page=4,width=0.40\textwidth]{plots/accuracy.pdf}
\caption{Top panels: comparison of differential distributions derived from our tables with three prior independent direct calculations. See sec.~\ref{sec:quality} for details.}
\label{fig:accuracy}
\end{figure}

To check the accuracy of the distributions obtained from the NNLO {\sc fastNLO} tables we compare these distributions with statistically independent NNLO calculations of the same differential distributions which were published in ref.~\cite{Czakon:2016dgf} for three different pdf sets: NNPDF30, CT14 and MMHT14 \cite{Harland-Lang:2014zoa} (all three pdf sets have $\alpha_s(m_Z) = 0.118$). The comparisons are shown in the upper panels of fig.~\ref{fig:accuracy}. The relative difference between the two calculations is small; for all four distributions and all three pdf sets it does not exceed 0.6\% in any bin (and for almost all bins it is about half that value). As can be seen in the lower panels of fig.~\ref{fig:accuracy}, this difference is comparable to the relative MC error of either the corresponding prior direct calculation or the correlated with the table NNPDF30-based direct calculation, discussed above in the context of the interpolation error estimate (recall that the dashed line in fig.~\ref{fig:accuracy} exactly corresponds to the blue line in fig.~\ref{fig:interpolation}).

As figs.~\ref{fig:interpolation},\ref{fig:accuracy} demonstrate, distributions derived from our {\sc fastNLO} tables are as accurate as the ones obtained from a direct calculation and can be readily used to obtain NNLO predictions with any pdf set without additional loss of numerical accuracy.

\section{Summary and Outlook}

In this work we produce {\sc fastNLO} tables for four top-quark pair differential distributions at NNLO corresponding to the ATLAS and CMS 8 TeV measurements \cite{Aad:2015mbv,Khachatryan:2015oqa}. The tables are publicly available and can be downloaded here \cite{web-tables}. These are the first publicly released {\sc fastNLO} tables at NNLO. The tables allow very fast calculation of these distributions with any pdf set and for different values of $\as$ through the LHAPDF interface. The tables will be indispensable in pdf fits as well as in any calculation of top-quark differential distributions with future pdf sets. We have verified the numerical accuracy of the NNLO differential distributions. It is high and comparable to all publicly available top-quark differential calculations. We intend to keep producing tables corresponding to other existing and future LHC measurements at 8 and 13 TeV. The most up-to-date set of released {\sc fastNLO} tables can be found at the website \cite{web-tables}.

\section{Acknowledgement}
The authors would like to thank Daniel Britzger and the {\sc fastNLO} collaboration for help with {\sc fastNLO} and useful discussion. The work of M.C. is supported in part by grants of the DFG and BMBF. The work of D.H. and A.M. is supported by the UK STFC grants ST/L002760/1 and ST/K004883/1. A.M. is also supported by the European Research Council Consolidator Grant ``NNLOforLHC2".

\appendix
\section{Using the tables}\label{sec:appendix}

To obtain the full NNLO differential cross section the tables need to be convoluted with a pdf set. For this purpose, a version of the {\sc fastNLO} toolkit is required. The tables have been tested for the latest public version (Version 2.3 pre-2212) which can be found on the {\sc fastNLO} website~\cite{fastNLO}. Here is a command line example for convoluting the $\Mtt$ table with the NNPDF30 pdf set through the LHAPDF~\cite{Buckley:2014ana} interface:
\begin{lstlisting}
fnlo-tk-cppread LHC8-Mtt-HT4-173_3-bin1.tab NNPDF30_nnlo_as_0118 1 LHAPDF no
\end{lstlisting}
The outputted cross-section for each bin, in pb/GeV, reads
\begin{lstlisting}
--------------------------------------------------------
LO cross section   NLO cross section  NNLO cross section 
--------------------------------------------------------
7.38589896926E-01  1.00686397386E+00  1.08054272971E+00
7.76050226541E-01  1.01684892793E+00  1.06831476452E+00  
4.72181816638E-01  6.14208165337E-01  6.51937131158E-01  
2.37714769748E-01  3.09500005873E-01  3.32361059337E-01  
9.50531653713E-02  1.24354657190E-01  1.34756258089E-01 
2.30309358260E-02  3.02533902041E-02  3.31987429211E-02  
2.66871047208E-03  3.51131256484E-03  3.94448314912E-03
\end{lstlisting}

\end{document}